\documentclass{ws-ijmpe}

\newcommand{\be}{\begin{equation}}
\newcommand{\ee}{\end{equation}}
\newcommand{\beq}{\begin{eqnarray}}
\newcommand{\eeq}{\end{eqnarray}}
\newcommand{\ba}{\begin{array}}
\newcommand{\ea}{\end{array}}

\begin{document}

\markboth{W. Satu{\l}a, J. Dobaczewski, W. Nazarewicz, M. Borucki, M. Rafalski}
{Isospin mixing in the vicinity of the $N=Z$ line}

\catchline{}{}{}{}{}

\title{Isospin mixing in the vicinity of the $N=Z$ line}

\author{W. Satu{\l}a$^a$, J. Dobaczewski$^{a,b}$, W. Nazarewicz$^{c,d,a}$,
M. Borucki$^e$, and  M. Rafalski$^a$}
\address{
$^a$Institute of Theoretical Physics, University of Warsaw,
         ul. Ho\.za 69, 00-681 Warsaw, Poland \\
$^b$Department of Physics, P.O. Box 35 (YFL),
University of Jyv\"askyl\"a, FI-40014 Finland \\
$^c$Department of Physics \& Astronomy,
  University of Tennessee, Knoxville, Tennessee 37996 \\
$^d$Physics Division, Oak Ridge National Laboratory, P.O. Box
  2008, Oak Ridge, Tennessee 37831 \\
$^e$Department of Physics, University of Warsaw,
    ul. Ho\.za 69, 00-681 Warsaw, Poland}

\maketitle

\begin{history}
\received{(received date)}
\revised{(revised date)}
\end{history}

\begin{abstract}
We present the isospin- and angular-momentum-projected nuclear density functional theory (DFT) and its applications to the isospin-breaking
corrections to the superallowed $\beta$-decay rates
in the vicinity of the $N=Z$ line.
A preliminary value obtained for the Cabbibo-Kobayashi-Maskawa
matrix element, $|V_{ud}|=  0.97463(24)$, agrees well with the
recent estimate by Towner and Hardy [Phys. Rev. C{\bf 77}, 025501 (2008)].
We also discuss new opportunities to study the symmetry
energy by using the isospin-projected  DFT.
\end{abstract}

\section{Introduction}

\noindent

The isospin-symmetry violation in atomic nuclei is predominantly
due to the Coulomb interaction that exerts long-range polarizations
on neutron and proton states. To consistently take into account
this polarization, one needs to
employ huge configuration  spaces. For that reason, an accurate description of isospin impurities in atomic nuclei, which
is  strongly motivated by the recent high-precision measurements of the $0^+
\rightarrow 0^+$ Fermi superallowed $\beta$-decay rates, is difficult to
be obtained in  shell-model
approaches, and  specific approximate methods are required.\cite{[Orm95a],[Tow08]}

The long-range polarization effects can be included within the
self-consistent mean-field (MF) or DFT
approaches, which are practically the only microscopic frameworks
available for heavy, open-shell
nuclei with many valence particles. These approaches, however, apart from
the  {\it physical\/} contribution to the isospin mixing, mostly caused by
the Coulomb field and, to a much lesser extent, by
isospin-non-invariant components of the nucleon-nucleon force,  also introduce the spurious isospin mixing  due to the {\it spontaneous\/} isospin-symmetry breaking.\cite{[Eng70],[Cau80],[Cau82]}

Hereby, we present results on the isospin mixing and
isospin symmetry-breaking corrections to the superallowed Fermi $\beta$-decay
obtained by using the newly developed isospin- and angular-momentum-projected DFT
approach without pairing.\cite{[Sat09],[Sat09a],[Sat10],[Sat10a]}
The model employs symmetry-restoration techniques to remove
the spurious isospin components and restore angular momentum symmetry, and
takes advantage of the natural ability of  MF to describe self-consistently
the subtle balance between the Coulomb force making  proton and
neutron wave functions different and the isoscalar part of the strong
interaction producing the opposite effect.

The paper is organized as follows. In Sec.~\ref{theo}, we describe the main theoretical
building blocks of the isospin- and angular-momentum-projected DFT.
Section~\ref{isomix} presents some preliminary applications of the formalism to the isospin symmetry-breaking corrections  to the Fermi superallowed
$\beta$-decay matrix elements, whereas Sec.~\ref{symm} discusses applications of the
isospin-projected DFT to nuclear symmetry energy. The summary is contained in  Sec.~\ref{summary}.

\section{The projected DFT framework}
\label{theo}

The building block of the isospin-projected DFT  is the  Slater determinant,
$|\Phi\rangle$, representing the self-consistent Skyrme-HF solution provided
by the HF solver HFODD.\cite{[Dob09d]} Self-consistency ensures that
the balance between the long-range Coulomb force and short-range strong
interaction, represented in our model by the Skyrme energy density functional (EDF), are properly taken
into account. The unphysical isospin mixing is taken care of by the
rediagonalization of the entire Hamiltonian in the good isospin basis, $|T,T_z\rangle$,
as described in Refs.\cite{[Sat10],[Sat10a]}
This yields the eigenstates:
\begin{equation}\label{mix2}
|n,T_z\rangle
= \sum_{T\geq |T_z|}a^n_{T,T_z}|T,T_z\rangle
\end{equation}
numbered by an index $n$.  The so-called isospin-mixing
coefficients (or, equivalently, isospin impurities)
are defined for the $n-$th eigenstate as
\begin{equation}
\alpha_C^n = 1 - |a^n_{T,T_z}|_{\text{max}}^2 ,
\end{equation}
where $|a^n_{T,T_z}|_{\text{max}}^2$ stands for the dominant amplitude in the wave function
$|n,T_z\rangle$.

Within the isospin- and angular-momentum-projected DFT, we
use the normalized basis of states $|I,M,K; T,T_z\rangle$
having both good angular momentum and good
isospin.\cite{[RS80]}
Here, $M$ and $K$ denote the angular-mo\-men\-tum components
along the laboratory and intrinsic $z$-axes, respectively. The  $K$ quantum
number is not conserved. In order to avoid problems with overcompleteness of
the basis, the $K$-mixing is performed by rediagonalizing
the Hamiltonian in the so-called {\it collective space}, spanned for each $I$
and $T$ by the {\it natural states\/}, $|IM;TT_z\rangle^{(i)}$, as described
in Refs.\cite{[Dob09d],[Zdu07a]} Such a rediagonalization yields the
eigenstates:
\begin{equation}   \label{KTmix}
|n; IM; T_z\rangle =
\sum_{i,T\geq |T_z|}
   a^{(n)}_{iIT} | IM; TT_z\rangle^{(i)} ,
\end{equation}
which are labeled by the index $n$ and by the conserved quantum numbers $I$, $M$, and
$T_z=(N-Z)/2$ [compare Eq.~(\ref{mix2})].

\begin{figure}\begin{center}
\includegraphics[angle=0,width=0.54\textwidth,clip]{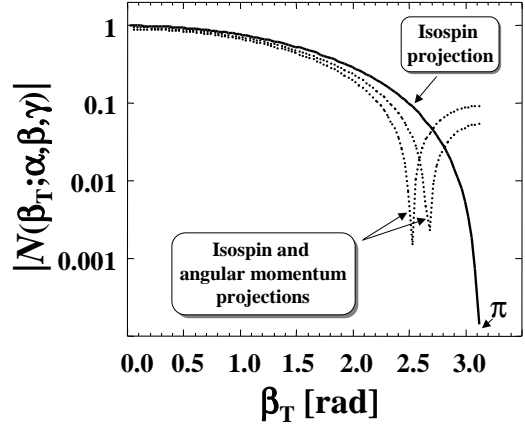}
\caption[T]{\label{fig1}
The absolute values of the norm kernels, $|{\cal N}(\beta_T; \alpha, \beta, \gamma )|
= |\langle \Phi | \hat{R}(\beta_T ) \hat{R}(\alpha, \beta, \gamma )
|\Phi\rangle|$, for a state in $^{14}$N calculated with the SLy4 EDF, plotted versus the rotation angle in the isospace $\beta_T$.
The solid curve, exhibiting the single singularity at $\beta_T = \pi $, corresponds to
the pure isospin-projected DFT theory, which is regular for all
Skyrme-type functionals.\protect\cite{[Sat10]}
The dotted lines correspond to two fixed sets of the Euler
angles in space, with $\alpha =\gamma \approx 0.314$, and
$\beta \approx 0.229$ (left curve)
and  $\beta \approx 1.414$ (right curve). The poles that appear
inside the integration region, $0<\beta_T<\pi$, give rise to singularities in
 the projected DFT approach.}
\end{center}\end{figure}

The isospin projection does not produce singularities in energy kernels; hence, it can be safely used with all commonly used EDFs.\cite{[Sat10]} Coupling the isospin and angular-momentum
projections, however, leads to singularities in both the norm (see
Fig.~\ref{fig1}) and energy kernels. This fact narrows the
applicability of the model to Hamiltonian-driven EDFs which,
for  Skyrme-type functionals, leaves only one option: the  SV
parametrization.\cite{[Bei75]} The alternative would be to use an
appropriate regularization scheme, which is currently under
development.\cite{[Lac09],[BD10]}

\section{Isospin-mixing and isospin-breaking corrections to superallowed
$\beta$-decay}
\label{isomix}

\begin{figure}\begin{center}
\includegraphics[angle=0,width=0.54\textwidth,clip]{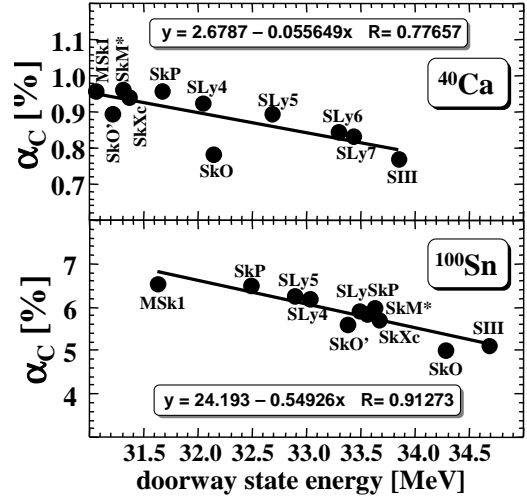}
\caption[T]{\label{fig2}
Isospin impurities in the ground states of $^{40}$Ca (upper panel) and
$^{100}$Sn (lower panel), plotted as functions of the excitation energy of
the doorway state for a set of commonly used Skyrme EDFs.\cite{[Ben03]} Results of the linear fits and the
corresponding regression
coefficients, $R$, are also shown.}
\end{center}\end{figure}

Evaluation of  $\alpha_C$ is
a prerequisite to calculate isospin corrections to reaction  and decay rates.
As is well known,\cite{[Aue83]}  isospin impurities are
the largest in $N=Z$ nuclei, increase along the $N=Z$ line with increasing
proton number, and are strongly quenched with increasing $|T_z|=|N-Z|/2$.
Such characteristics were also early estimated based on the perturbation
theory\cite{[Sli65]} or hydrodynamical model.\cite{[Boh67]} Quantitatively,
after getting rid of the spurious mixing, which lowers the true $\alpha_C$ by as
much as 30\%,\cite{[Sat09a]}
the isospin impurity increases from a fraction of a percent in very light
$N=Z$ nuclei to  $\sim$0.9\%  in $^{40}$Ca, and $\sim$6.0\% in $^{100}$Sn,
as shown in Fig.~\ref{fig2}. In the particular case of $^{80}$Zr, the
calculated impurity of 4.4\% agrees well with the empirical value  deduced from
the giant dipole resonance $\gamma$-decay studies.\cite{[Cam10a]} This  makes us believe that our model  is indeed capable of
capturing essential physics associated with the isospin mixing. Unfortunately,
current experimental errors are too large to discriminate between different
parametrizations of the Skyrme functional. The variations between EDFs in Fig.~\ref{fig2} result in $\sim$10\% uncertainty in calculated
values of $\alpha_C$.

The magnitude of theoretical $\alpha_C$ is quite well correlated
with the excitation energy, $E_{T=1}$, of the $T=1$ doorway state,
see Fig.~\ref{fig2}. However, in order to make a precise determination of $E_{T=1}$, spectroscopic quality EDFs are needed, and this  is not yet the case.\cite{[Kor08]} This explains why the values of $\alpha_C$
do not  correlate well with basic EDF characteristics, including the isovector and isoscalar effective mass,
symmetry energy, binding energy per particle, and
incompressibility (see discussion in Ref.\cite{[Sat10a]}).

\begin{figure}\begin{center}
\includegraphics[angle=0,width=0.54\textwidth,clip]{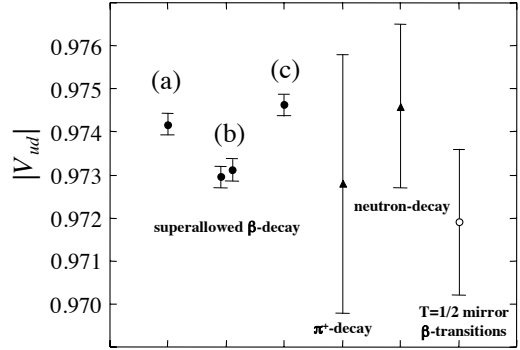}
\caption[T]{\label{fig3}
Values of $|V_{ud}|$ deduced from the superallowed $\beta$-decay
(full circles) for three different sets of the $\delta_C$ corrections calculated in:
Ref.\protect\cite{[Tow08]} (a);  Ref.\protect\cite{[Lia09]} with NL3 and
DD-ME2 Lagrangians (b); and in the present work (c).
Triangles mark values of $|V_{ud}|$ obtained from the
pion-decay\protect\cite{[Poc04]} and neutron-decay\protect\cite{[Ams08]} studies,
respectively. The open circle shows the value deduced from the
$\beta$-transitions in $T=1/2$ mirror nuclei.\protect\cite{[Nav09a]}
}
\end{center}\end{figure}

Increasing demand on precise values of  isospin impurities has been
stimulated by the recent high-precision measurements of superallowed
$\beta$-decay rates.\cite{[Har05c],[Tow08]}
A reliable determination of the corresponding
isospin-breaking correction, $\delta_C$,
requires the isospin- and angular-momentum-projected DFT.\cite{[Sat10a]}.
This correction is obtained by calculating
the $0^+  \rightarrow 0^+$ Fermi
matrix element of the isospin raising/lowering operator $\hat T_{\pm}$ between
the ground state (g.s.) of the even-even nucleus  $| I=0, T\approx 1, T_z = \pm 1 \rangle$
and its isospin-analogue partner in the $N=Z$ odd-odd nucleus, $|I=0, T\approx
1, T_z = 0 \rangle$:
\begin{equation}\label{fermime}
|\langle I=0, T\approx 1,
T_z = \pm 1 | \hat T_{\pm} | I=0, T\approx 1, T_z = 0 \rangle |^2 \equiv 2 (
1-\delta_C ).
\end{equation}

To determine the  $|I=0, T\approx 1, T_z = 0 \rangle$ state
in the odd-odd $N=Z$ nucleus, we
first compute the so-called
antialigned g.s.\ configuration, $|\bar \nu \otimes \pi \rangle$ (or  $| \nu
\otimes \bar \pi \rangle$), by placing the odd neutron and the odd proton in
the lowest available time-reversed (or signature-reversed)
HF orbits.
Then, to correct for the fact that the antialigned
configurations manifestly break the isospin symmetry,\cite{[Sat10]} that is,
$|\bar \nu \otimes \pi \rangle \approx \frac{1}{\sqrt 2} (|T=0 \rangle + |T=1
\rangle )$, we apply the isospin and angular-momentum projections to create
the  basis $|I,M,K,T,T_z=0 \rangle$, in which the total Hamiltonian is rediagonalized (see  Sec.~\ref{theo}).
A similar scheme is used to compute the
$| I=0, T\approx 1, T_z = \pm 1 \rangle$ states in the even-even nuclei.

Our studies indicate\cite{[Sat10a]} that to obtain a fair estimate
of $\delta_C$ for $A<40$ and $A>40$ nuclei, one needs to use large
harmonic oscillator bases consisting of at least $N=10$ and 12
full shells, respectively. Even then, the results
are subject to systematic errors due to the basis cut-off, which can be
estimated to be $\sim$10\%.
Despite the fact that  not all $N=12$ calculations in heavy ($A> 40$) nuclei
have yet been completed, and that owing to the
 shape-coexistence effects, there are still some
ambiguities concerning the global minima, our  preliminary results point
to  encouraging conclusions. Namely,  the mean value of the structure-independent
statistical-rate function $\bar{{\cal F}}t$,\cite{[Har05c]} obtained for 12 out of
13 transitions known empirically with high precision (excluding the
$^{38}$K$\rightarrow$$^{38}$Ar case), equals $\bar{{\cal F}}t = 3069.4(10)$,
which gives the value of the CKM matrix element equal to $|V_{ud}| =  0.97463(24) $.
These values match well those obtained by Towner and Hardy in their
recent compilation\cite{[Tow08]} (see Fig.~\ref{fig3}).
Because of a poor spectroscopic quality of the SV parameterization, the confidence
level\cite{[Tow10]} of our results is poor. Nevertheless, it should be
stressed that our method is quantum-mechanically consistent (see
discussion in Refs.\cite{[Mil08],[Mil09]}) and contains no adjustable free parameters.

\section{Symmetry energy}\label{symm}

\begin{figure}\begin{center}
\includegraphics[angle=0,width=0.46\textwidth,clip]{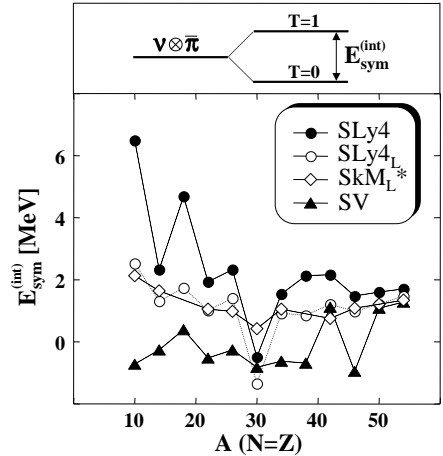}
\caption[T]{\label{fig4}
Top: schematic illustration of the isospin-symmetry-breaking
mechanism in MF of odd-odd $N=Z$  nuclei. Bottom:
 $E_{\text{sym}}^{\text{(int)}}$ in odd-odd $N=Z$ nuclei calculated with
SLy4, SV, SLy4$_L$, and SkM$^*_L$ EDFs. See text for details.}
\end{center}\end{figure}

The spontaneous violation of isospin symmetry in all but isoscalar MF configurations of
$N=Z$ nuclei offers a way to study the nuclear symmetry
energy.  The idea, which is schematically
sketched in the upper portion of Fig.~\ref{fig4}, invokes  the mixed-symmetry
antialigned $|\bar\nu \otimes \pi\rangle$ (or $|\nu \otimes \bar\pi\rangle$)
configuration in an odd-odd $N=Z$ nucleus. By applying the isospin projection to the
HF state $|\bar\nu \otimes \pi\rangle$, one decomposes  it into the isoscalar $T=0$ and isovector
$T=1$ parts. As argued below, the magnitude of the splitting, $E_{\text{sym}}^{\text{(int)}}$,
depends on the isovector channel of a given EDF, i.e., its symmetry energy.

For the Skyrme-type EDFs, the  symmetry energy
in the nuclear matter limit can be decomposed as:\cite{[Sat06w2]}
\begin{equation}\label{nmsym}
a_{\text{sym}} = \frac{1}{8}\varepsilon_{FG} \left( \frac{m}{m_0^\star}
\right) +
\left[ \left( \frac{3\pi^2}{2}\right)^{2/3} C_1^\tau \rho^{5/3}
+ C_1^\rho \rho \right]
\equiv  a_{\text{sym}}^{\text{(kin)}} + a_{\text{sym}}^{\text{(int)}}.
\end{equation}
The first term in Eq.~(\ref{nmsym}) is associated with the isoscalar part of the nucleon-nucleon interaction
and primarily depends on the mean single-particle level spacing at the Fermi energy. This term
is scaled by the inverse isoscalar effective mass. The second
(interaction) term, is related to the isovector part of the Skyrme-EDF:
$\delta {\cal H}_{t=1} = C_1^\rho \rho_1^2 + C_1^\tau \rho_1 \tau_1$
(for definitions, see Ref.\cite{[Ben03]} and references quoted therein).

The value of $E_{\text{sym}}^{\text{(int)}}$ appears to be mainly sensitive to the
interaction term, which is illustrated in Fig.~\ref{fig4}.
Indeed, despite the fact that
SLy4 and SV EDFs have similar values of  $a_{\text{sym}}$ (equal
to 32\,MeV and 32.8\,MeV, respectively), the
corresponding energy splittings $E_{\text{sym}}^{\text{(int)}}$ differ substantially.
The reduced values of $| E_{\text{sym}}^{\text{(int)}} |$ in  SV
are due to its small value of $a_{\text{sym}}^{\text{(int)}} = 1.4 $\,MeV,\footnote{
This small value shows how unphysical are the consequences of
the saturation mechanism built into  SV through the strong
momentum dependence and results in an unphysically low
 isoscalar effective mass  $m^*/m\approx 0.38$. Although  SV  has
a relatively reasonable global strength of the symmetry energy $a_{\text{sym}}$, its physical origin is incorrect.}
 which is
an order of magnitude smaller than the corresponding SLy4 value:
 $a_{\text{sym}}^{\text{(int)}} = 14.4$\,MeV.

An interesting aspect of  our analysis of
$E_{\text{sym}}^{\text{(int)}}$ relates to its dependence on the
time-odd terms, which  are poorly constrained for Skyrme EDFs.
To quantify this dependence, we have performed
calculations by using the SLy4$_L$ and SkM$^*_L$ functionals, which
have the spin coupling constants adjusted to the Landau parameters.\cite{[Ben02],[Zdu05]}
These EDFs have different  values of $a_{\text{sym}}$ but the same
$a_{\text{sym}}^{\text{(int)}} = 14.4$\,MeV. The similarity of the calculated energy splittings shown in Fig.~\ref{fig4} confirms that this quantity
primarily depends on the isovector terms of the functional. Moreover, its
significant dependence on the time-odd terms opens up new options for adjusting
the corresponding coupling constants to experimental data. This will certainly require the simultaneous restoration of isospin and
angular-momentum symmetries, as presented in this study.

\section{Summary}\label{summary}

In summary, the isospin- and angular-momentum-projected DFT
calculations have been performed to estimate  the isospin-breaking
corrections to $0^+ \rightarrow 0^+$ Fermi superallowed $\beta$-decays.
Preliminary results for the average value of the nucleus-independent $\bar{\cal
F}t = 3069.4(10)$ and the amplitude $|V_{ud}| = 0.97463(24)  $ were found
to be  consistent with the recent estimates by  Towner and Hardy,\cite{[Tow08]}
notwithstanding a  low spectroscopic quality of the Skyrme EDF SV used.

Applicability of the isospin-projected DFT  to analyze
the nuclear symmetry energy has also been discussed. It has been demonstrated
that the isospin projection offers a  rather unique opportunity to study the
interaction part of the symmetry energy
in the odd-odd $N=Z$ nuclei and that this quantity is influenced
by time-odd fields of the energy density functional.

This work was supported in part by the Polish Ministry of Science
under Contract Nos.~N~N202~328234 and N~N202~239037, Academy of Finland and
University of Jyv\"askyl\"a within the FIDIPRO programme, and by the Office of
Nuclear Physics,  U.S. Department of Energy under Contract Nos.
DE-FG02-96ER40963 (University of Tennessee) and
DE-FC02-09ER41583  (UNEDF SciDAC Collaboration).
We acknowledge the CSC - IT Center for Science Ltd, Finland for the
allocation of computational resources.


\end{document}